\documentclass{cpbtex}

\usepackage{graphicx}
\usepackage{caption}
\usepackage{ulem}
\usepackage{color}
\begin{document}
\begin{CJK*}{GBK}{song}

\title{Unveiling the early stage evolution of local atomic structures in the crystallization process of a metallic glass}


\author{Lin Ma \begin{CJK}{UTF8}{gbsn}(马琳)\end{CJK}$^{1}$\thanks{These two authors contribute equally to this work.},
	 \ Xiao-Dong Yang \begin{CJK}{UTF8}{gbsn}(杨晓东)\end{CJK}$^{1}$\footnotemark[1],
	 \ Feng Yang \begin{CJK}{UTF8}{gbsn}(杨锋)\end{CJK}$^{1}$,\\
     \ Xin-Jia Zhou \begin{CJK}{UTF8}{gbsn}(周鑫嘉)\end{CJK}$^{1}$,
	 \ Zhen-Wei Wu \begin{CJK}{UTF8}{gbsn}(武振伟)\end{CJK}$^{1}$\thanks{Corresponding author Email: zwwu@bnu.edu.cn}\\
$^{1}${Institute of Nonequilibrium Systems, School of Systems Science, Beijing Normal University,}\\
{ 100875 Beijing, China}} 

\date{\today}
\maketitle

\begin{abstract}
The early stage evolution of local atomic structures in a multicomponent metallic glass during its crystallization process has been investigated via molecular dynamics simulation. It is found that the initial thermal stability and earliest stage evolution of the local atomic clusters show no strong correlation with their initial short-range orders, and this leads to an observation of a novel symmetry convergence phenomenon, which can be understood as an atomic structure manifestation of the ergodicity. Furthermore, in our system we have quantitatively proved that the crucial factor for the thermal stability against crystallization exhibited by the metallic glass, is not the total amount of icosahedral clusters, but the degree of global connectivity among them.
\end{abstract}



\section{Introduction}
Since the advents of multicomponent metallic glasses (MGs) in the early 1990s, the initial atomic-scale structure evolution and nanocrystallization under annealing or mechanical loading treatments of their glass-forming liquids, have been extensively investigated for both fundamental and practical reasons~\ucite{Wang2019Dynamic, Sheng2006Atomic,Cheng2009Atomic,Pauly2010Transformation,Inoue2000Stabilization,Wang2011Atomic,Liu2008Atomistic,Wang2003In,Luo2004Icosahedral,Chen2006Molecular,Martin2004Nanocrystallization,DellaValle1994Microstructural,Kelton2003First,Cheng2011Atomic,Li2009Structural, Wu2013Effect, Jiang2016Effect, Mokshin2008Shear, Mokshin2010Crystal,Wu2021Quantitative,Cui2022Spatial}.  Various unique internal structures of MGs underlie their interesting properties, which render MGs potentially useful for distinct applications\ucite{Liu2008Atomic,Gao2021Hydrogen,Ye2016Generalized,Chen2014Atomic}.  For the excellent mechanical properties of glasses with nanocrystalline inclusions and as well fundamental meanings to solid state physics, numerous studies have been conducted to comprehend how amorphous alloys crystallize at nanoscales. However, due to the lack of directly time-resolved data of atomic structure evolutions at the early stage of crystallization, clear pictures of how localized atomic  structures rearranging themselves, developing original translational symmetry, and eventually forming nanocrystals with three-dimensional (3-D) periodicity, remain less well-understood at present. Especially the questions that which local parts of the MGs will the firstly get into the structure with embryonic translational symmetry, what kind of features these local parts have, and how they will evolve in the future, are still mysterious at present.

Through high-resolution transmission electron microscopy (HRTEM) and advanced structural analysis techniques, researchers have accumulated massive precious experimental data to study and reveal the MGs' structure evolution under certain treatments\ucite{Pauly2010Transformation,Inoue2000Stabilization,Wang2011Atomic,Chen2006Molecular,Martin2004Nanocrystallization}.  A series of meaningful achievements have been obtained.  But HRTEM cannot give us the answers of these problems mentioned above.  A clearly physical picture that can describe the details of the evolutions associated with different atomic-scale structures and the features of these localized structures has not been obtained now.  The quantitative description of this process is still in its early stages.

The icosahedral short-range order (ISRO) in MGs is generally adopted as a structural indicator for the fundamental process underlying structural relaxation\ucite{Cheng2008Indicators,Wu2013Correlation,Li2017Five,Qiao2023A}. However, in some MGs the ISRO is absent. As shown in Ref.~\cite{Peng2011Structural}, atomic symmetry is a more general concept in the glass-forming alloys in which both local fivefold and translational symmetries are present. Both experimental and simulated researches have shown that local clusters symmetry plays an important role in determining many properties of MGs. Therefore, it is worth to check that whether the degrees of the local fivefold symmetry (LFFS) and the local translational symmetry (LTS) connected with the analysis of local icosahedral clusters, especially the spatial topological order of them, are more crucial for the understanding of the structural evolution and thermal stability of MGs.

In this study, to address above challenges we performed molecular dynamics (MD) simulation with the LAMMPS code\ucite{Pli95}, and the degrees of LFFS and LTS were used as the structural indicators to identify and characterize the irreversible structural evolutions in the crystallization process. The spatial correlation of the icosahedral clusters in the glass structures and its influence on the thermal stability of MGs against crystallization has also been investigated. We find that the initial stability and the very beginning evolution of the local atomic structure are basically unaffected by the degree of LFFS of these regions. Moreover, it is found that not the quantity of the icosahedral clusters but the connectivity among them, will be the crucial factor for the thermal stability against crystallization in amorphous alloys.

\section{Methodology}
In our MD simulations, a model system of Zr$_{65}$Al$_{30}$Cu$_{5}$ containing $10000$ atoms in a cubic box with periodic boundary conditions was adopted.  The realistic embedded-atom method potential used is the one developed based on the data in Ref.~\cite{Zhou2004Misfit}. The integration time step of all simulations was set to $2$ fs.  In the process of sample preparation, it was first melted and equilibrated at $T=2500$K for $1$ ns, and then cooled down to $290$K with a cooling rate of $10^{12}$K/s ($10^{13}$K/s), during which the cell size was adjusted to maintain zero pressure in the NPT ensemble. Finally, the sample was relaxed at $290$K for another $1$ ns. To ensure the reliability of the MD-generated atomic configurations, the pair correlation functions (PCFs) of the model were calculated to declare the amorphous states of the samples. For investigating the crystallization process, annealing treatment was applied on the sample at $T=650$K (above the kinetic glass transition temperature $T_g=615$K) in the NPT ensemble for doing the following analysis.

\section{Results and Discussion}

\begin{figure}
	\begin{center}
		\includegraphics[width=0.39\columnwidth]{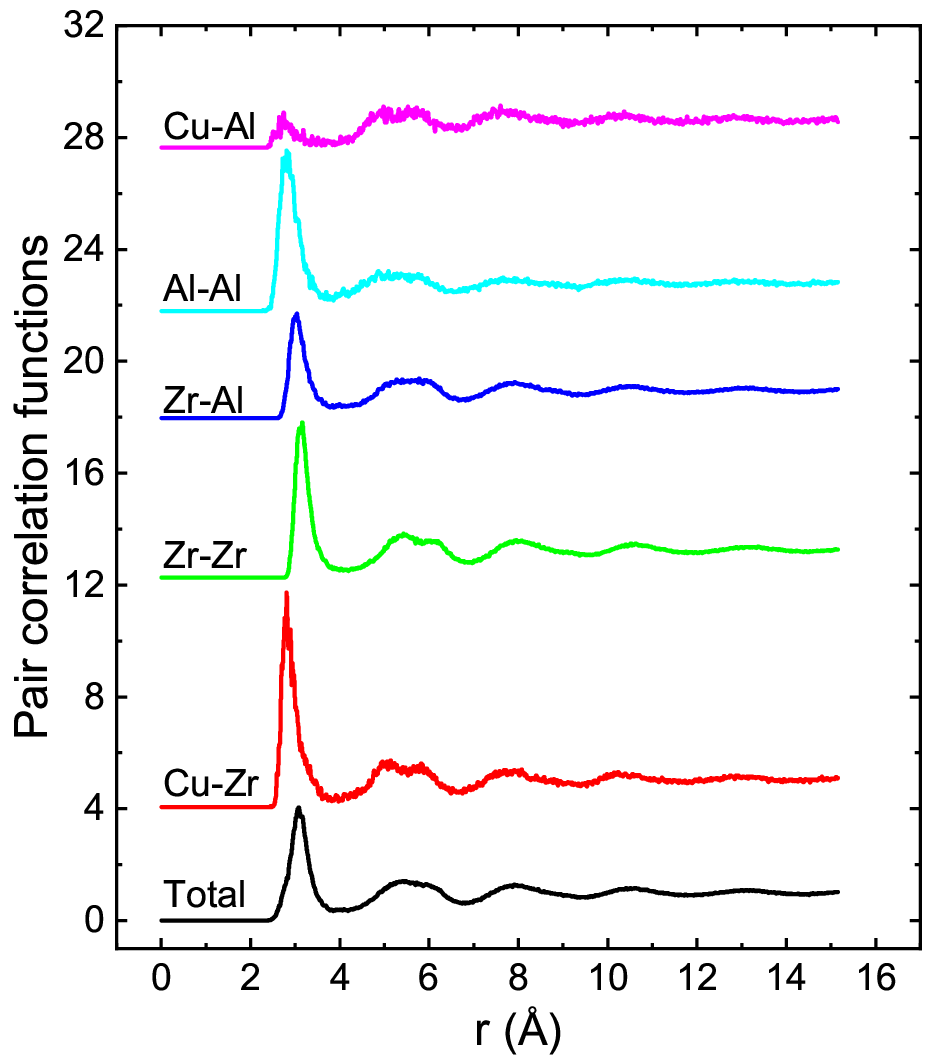}\\[5pt]  
		\caption{\parbox[c]{15.0cm}{\footnotesize{\bf Fig.~1.} (color online) The calculated PCFs of the MD-generated atomic configurations obtained at a cooling rate of $10^{13}$K/s. For clarity, the noisy and scattered Cu-Cu partial PCF is not included.}}
		\label{Fig:PCFs}
	\end{center}
\end{figure}

\begin{figure}
	\begin{center}
		\includegraphics[width=0.75\columnwidth]{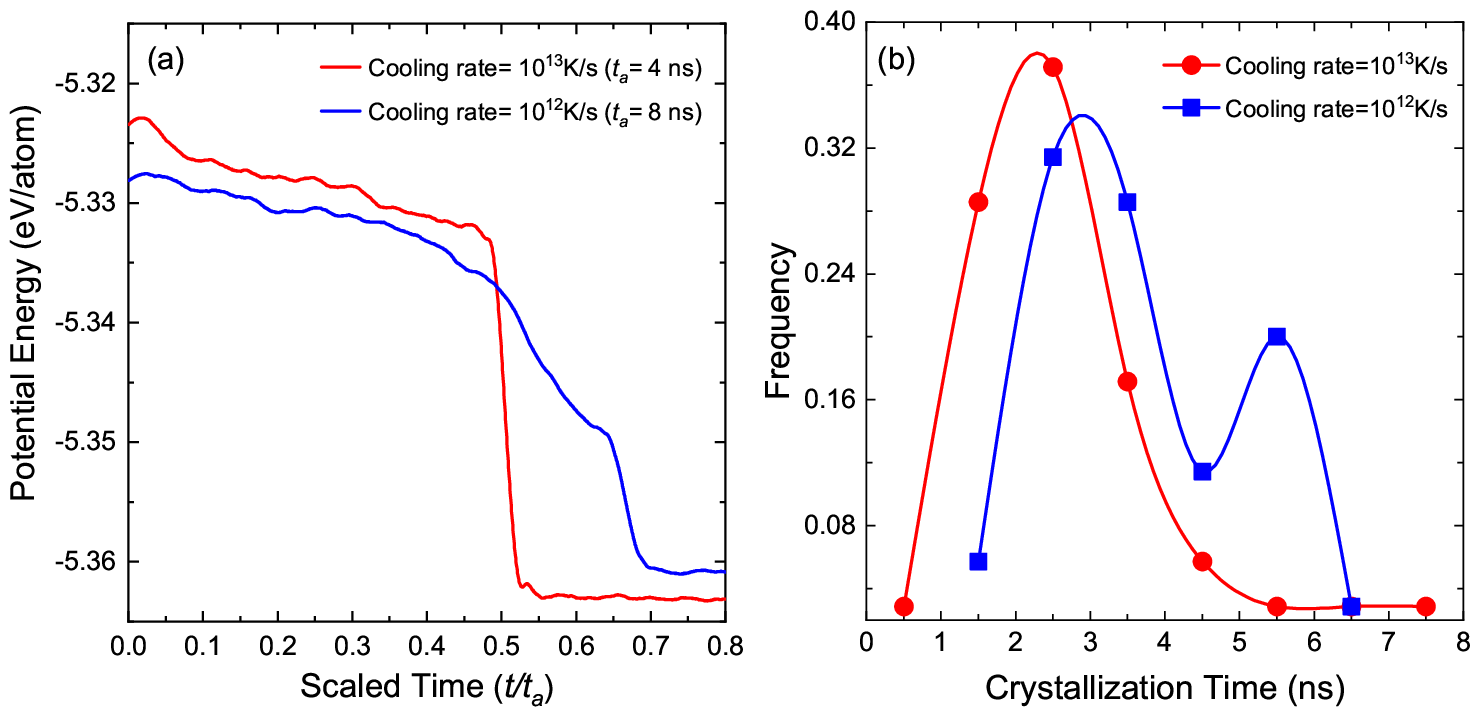}\\[5pt]  
		\caption{\parbox[c]{15.0cm}{\footnotesize{\bf Fig.~2.} (color online) (a) The annealing time dependence of the system potential energy during the isothermal annealing treatment at $650$K. For clarity, the time is scaled by $t_a$ which is the total annealing time for the two samples respectively. (b) The distribution of isoconfigurational averaged crystallization-time.}}
		\label{Fig:PE}
	\end{center}
\end{figure}

The PCFs of the amorphous Zr$_{65}$Al$_{30}$Cu$_5$ alloy obtained through MD simulation are presented in Fig.~\ref{Fig:PCFs}. The overall PCF exhibits a shoulder on the second peak, and the Zr-Zr partial PCF displays the largest distance for the first peak, consistent with the previous experimental findings on these alloys\ucite{Yang2006Design}. Additionally, the Cu-Zr and Zr-Zr partial PCFs show a characteristic splitting of the second peak into two subpeaks, which is a common feature observed for metallic glasses\ucite{Mendelev2007Using}. These results, to some extent, affirm the reliability of the MD-generated atomic configurations for the subsequent analysis. Figure~\ref{Fig:PE}(a) presents the potential energy (PE) evolution of the entire system as a function of scaled-time during the isothermal annealing process. The onset of crystallization is evident by the PEs exhibiting a drop at a scaled-time of $0.5 \sim 0.6$. This decrease signifies the initiation of crystallization. Notably, Figure~\ref{Fig:PE}(a) also illustrates distinct variation trends in the PEs of two samples generated with different cooling rates.
It is worth to note that we do not emphasise the relationship between the rate of cooling and the rate of crystallization here. The two different cooling rates were used to produce two different configurations for the subsequent analysis. In other words, the only purpose here choosing two slightly different cooling rates is to generate two samples that are similar in short-range orders but with a bit subtle difference in medium-range orders (as it can be seen below). And we also performed the iso-configurational runs to take the thermal fluctuation effect into account. It was found that only $\sim$8 cases out of 40 that the fast-quenched sample took much longer time to crystalize than the slow-quenched one, and the {\it simply} isoconfigurational averaged crystallization-time (in practical defined as the middle-term of the PE's drops) of the slow-quenched sample is $3.6$ns which is indeed statistically longer than the one of fast-quenched sample's $2.8$ns. More impressively, as it can be seen in Fig.~\ref{Fig:PE}(b), the distribution of the crystallization-time for the slow-quenched sample is bimodal, in which there is a peak (This peak remains after increasing the statistical sample size to 80, ruling out the possibility that this is a statistical rise and fall.) located at long-time intervals, what is significantly different from the one of the fast-quenched samples. Apparently, all above results indicate that the initial atomic structure indeed plays an important role on the thermal stability of our metallic liquid.
The underlying reasons for this intriguing phenomenon will be further explored in subsequent sections of this work.

\begin{figure}
	\begin{center}
		\includegraphics[width=0.7\columnwidth]{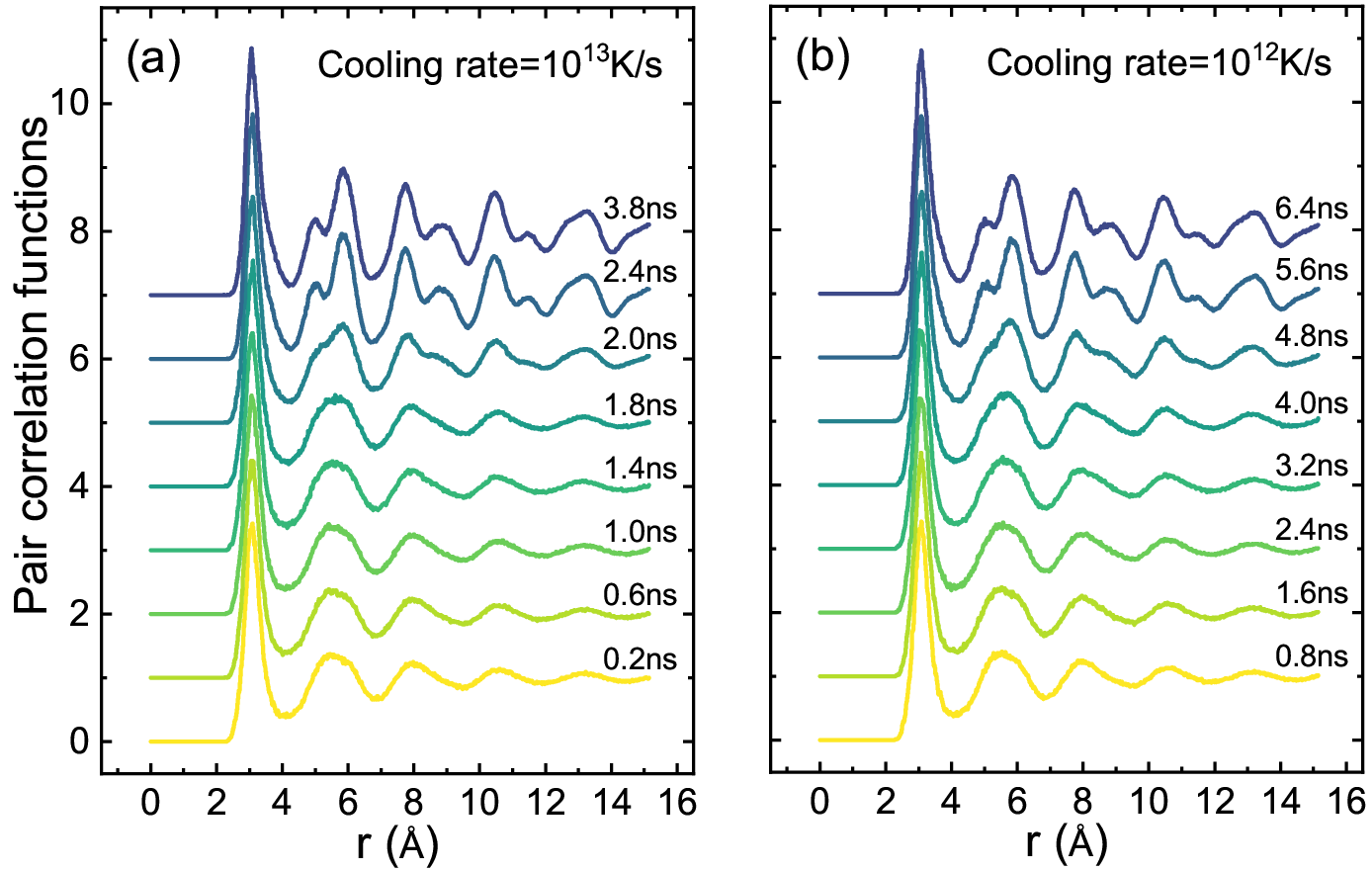}\\[5pt]  
		\caption{\parbox[c]{15.0cm}{\footnotesize{\bf Fig.~3.} (color online)
        Changes in the total PCFs during the crystallization of two MG samples obtained at different cooling rates. (a) Cooling rate = $10^{13}$ K/s. (b) Cooling rate = $10^{12}$ K/s.
        }}
		\label{Fig:PCFsPCFs}
	\end{center}
\end{figure}

\begin{figure}
	\begin{center}
		\includegraphics[width=0.6\columnwidth]{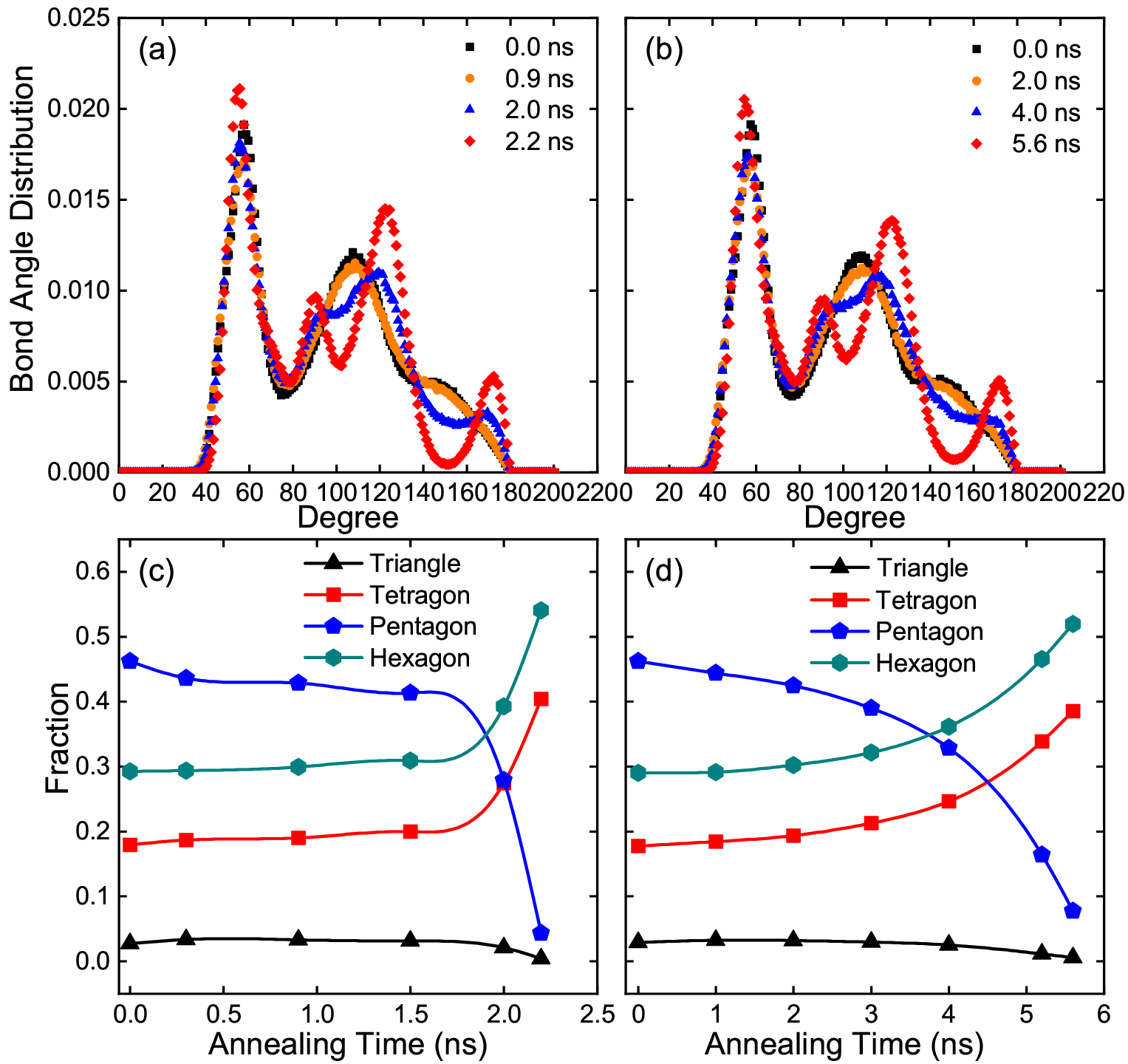}\\[5pt]  
		\caption{\parbox[c]{15.0cm}{\footnotesize{\bf Fig.~4.} (color online) The variations of the bond-angle distributions and degree of LFFS and LTS of the MGs system during the annealing process.  (Cooling rate=$10^{13}$K/s, in (a) and (c); cooling rate=$10^{12}$K/s, in (b) and (d).)}}
		\label{Fig:BondAngleFraction}
	\end{center}
\end{figure}

The atomic structure changes throughout the annealing process are indicated by the temporal evolution of the PEs, as demonstrated by the evolving PCFs in Fig.~\ref{Fig:PCFsPCFs}, which illustrate a transition from an amorphous state to a crystalline solid. To comprehensively investigate the local structural rearrangement preceding crystallization, we employed the Voronoi tessellation analysis\ucite{Finney1970Random,Borodin1999Local}. This method involves dividing the three-dimensional space into cells centered around each atom. A plane is drawn to bisect each line connecting the central atom and its nearest-neighbors, resulting in the formation of a Voronoi cell enclosed by these inner planes. The Voronoi index $<n_3, n_4, n_5, n_6>$ is utilized to characterize the atomic clusters surrounding each atom, where $n_i$ represents the number of $i$-edged faces of the corresponding Voronoi cell. The presence of an $i$-edged face reflects the local symmetry of the central atom with its nearest neighbors in a specific direction. Notably, faces with three, four, and six edges exhibit LTS features, while pentagon faces indicate the presence of LFFS. The degree of LFFS in local structures is quantified using a specific metric defined as $d_5=n_5 / \sum_{i} n_i$~\ucite{Peng2011Structural}. Additionally, by analyzing the coordination information of atoms obtained from the Voronoi analysis, the system's global bond-angle distributions can be measured. We selected a number of different annealing time periods and analyzed the local structures of the corresponding atomic configurations. The results shown in Fig.~\ref{Fig:BondAngleFraction} reveal the changes in bond angle distributions, which demonstrate a gradual increase in LTS features within the system throughout annealing. The initial bond angle distribution clearly indicates the amorphous state of the sample. As the annealing process progresses, characteristic angles of $90^{\circ}$, $120^{\circ}$, and $180^{\circ}$ gradually emerge and become more prominent, suggesting an enhancement of ordering in the system. The variation in the fraction of $i$-edged faces in the Voronoi polyhedrons (Fig.~\ref{Fig:BondAngleFraction}) also supports the aforementioned trend. The evolution of these local symmetries behaves similarly in the samples with different cooling rates, despite the fact that the sample generated with a lower cooling rate evolves more gradually while the one generated with a higher cooling rate crystallizes more suddenly after a bit gentle embryo state processing. Below we will discuss this behavior in more details.

\begin{figure}
	\begin{center}
		\includegraphics[width=0.7\columnwidth]{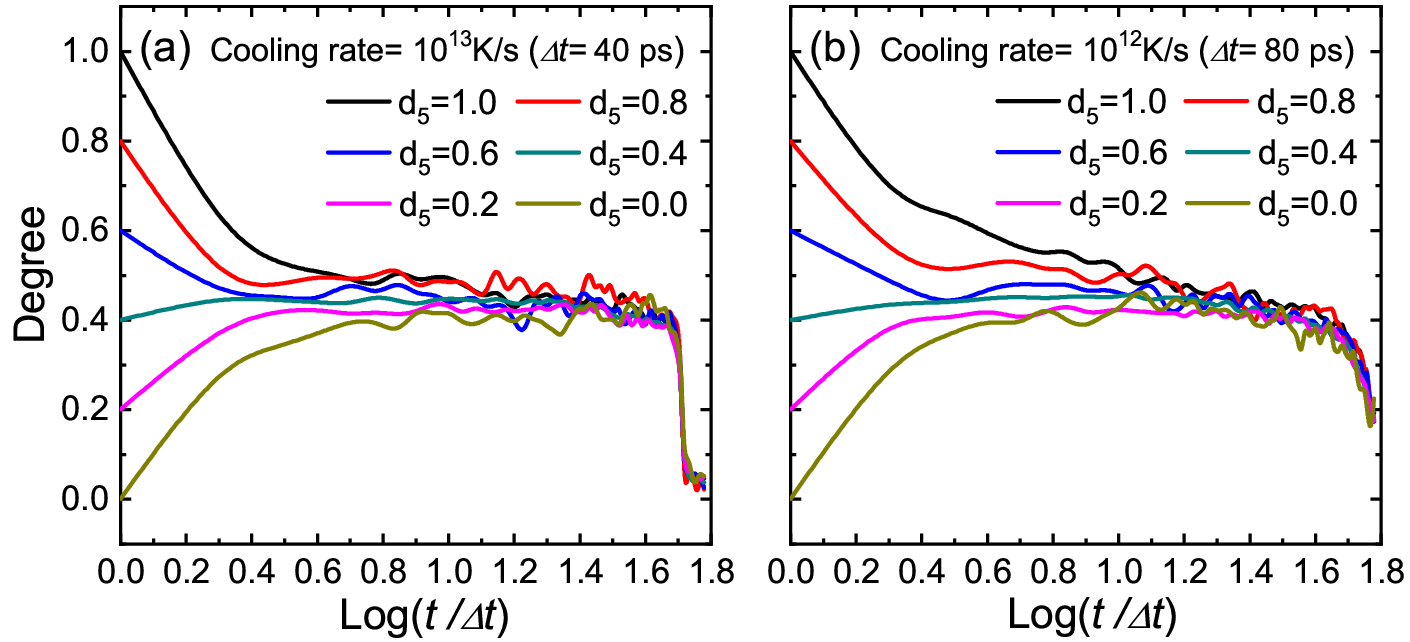}\\[5pt]  
		\caption{\parbox[c]{15.0cm}{\footnotesize{\bf Fig.~5.} (color online)  The evolution of $d_5$ of different local atomic regions classified by their corresponding original degree of $d_5$ (from $0.0$ to $1.0$) for the samples with two different cooling rate during the annealing process. (a) Cooling rate=$10^{13}$K/s. (b)Cooling rate=$10^{12}$K/s. $\Delta t$ is the configuration sampling time interval.}}
		\label{Fig:Fractiond5}
	\end{center}
\end{figure}

\begin{figure}
	\begin{center}
		\includegraphics[width=0.8\columnwidth]{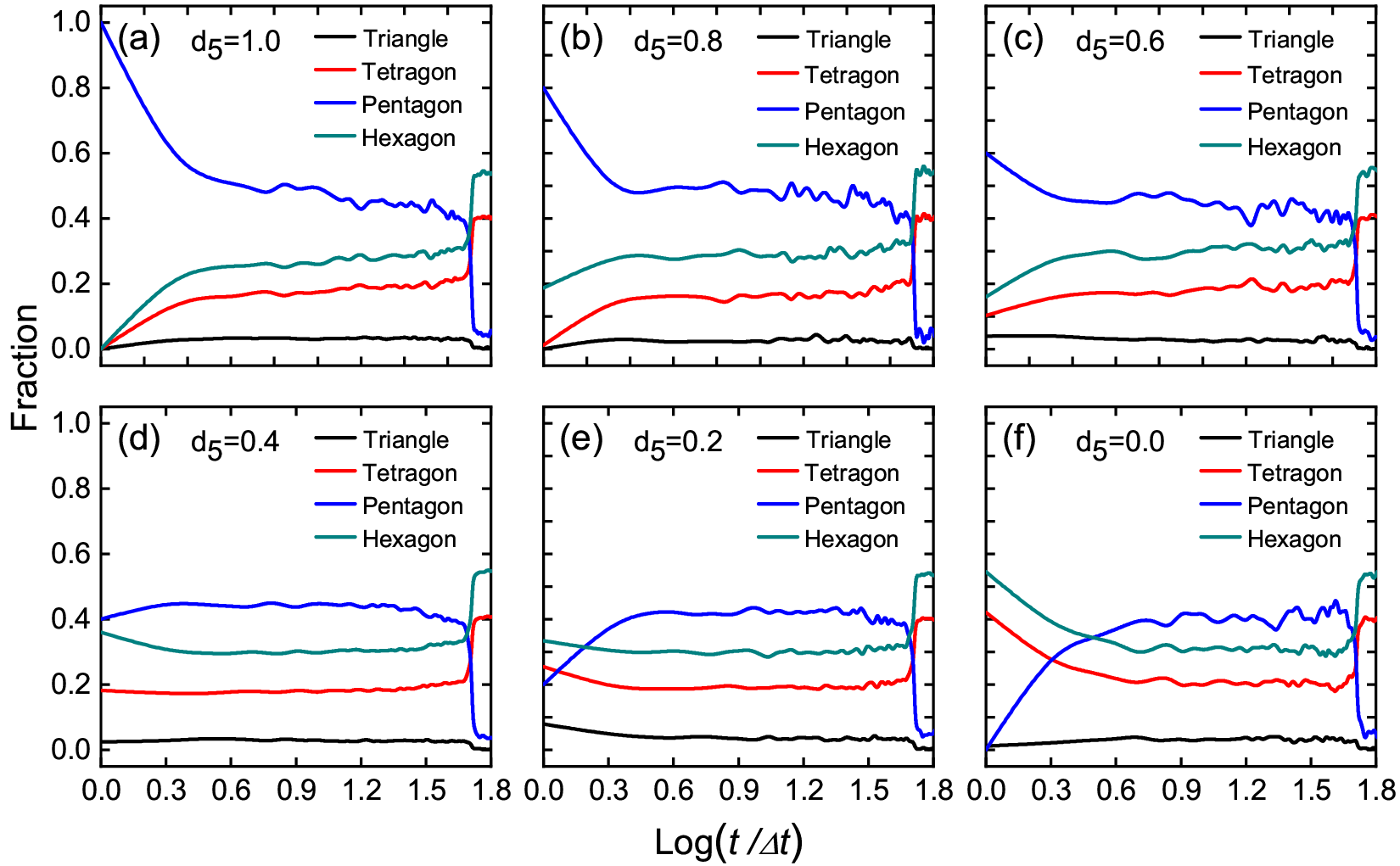}\\[5pt]  
		\caption{\parbox[c]{15.0cm}{\footnotesize{\bf Fig.~6.} (color online) The change of $d_3$, $d_4$, $d_6$, and $d_5$ of different local atomic regions classified by their corresponding original degree of $d_5$ ((a)$d_5=1.0$, (b) $d_5=0.8$, (c)$d_5=0.6$, (d)$d_5=0.4$, (e)$d_5=0.2$, (f)$d_5=0.0$.) for the sample with a cooling rate of $10^{13}$K/s during the annealing process.  The time interval $\Delta t$ on the x-axis corresponds to $40$ ps.}}
		\label{Fig:Fractiond3d4d5d6}
	\end{center}
\end{figure}

To provide a detailed description of the evolution of local structures with different symmetry characteristics, we used the degree of LFFS $d_5$ in Voronoi cells associated with local atomic clusters to label corresponding regions and then examine the subsequent variation of $d_5$ of these regions in the annealing process. As shown in Fig.~\ref{Fig:Fractiond5}, in contrast to some previous observations where regions with a higher degree of LFFS exhibited greater stability, and regions with lower degree of LFFS underwent more frequent irreversible rearrangements, our results indicate that the thermal stability and evolution speed of local structures are basically unaffected by the degree of $d_5$ of these regions during annealing. Notably, we observed a phenomenon of symmetry convergence (Fig.~\ref{Fig:Fractiond5}), where local structures with different initial LFFS characteristics undergo transformations in their localized environments, leading the entire system towards a state where the degree of LFFS is approximately 0.4.
After this convergence phenomenon, $d_5$ begins to fall fast, indicating that the system is crystallising, the growth and development of nanocrystals occur randomly within the samples. Based on the aforementioned discussion, our aim is to elucidate the key mechanism behind the nucleation of high-density nanosized crystals during the initial stages of crystallization in MGs under supercooled liquid conditions\ucite{Wang2011Atomic,Liu2008Atomistic,Schroers1999Continuous,Schroers2000Crystallization,Pauly2010Criteria}. We propose that the convergence of local structure symmetry results in the system transitioning into an embryo stage, a process that can be understood as the atomic-structure manifestation of ergodicity, which makes sure that every local region in the sample has the potential to nucleate and grow into crystals from a liquid state. To further substantiate our viewpoints, we provide a more detailed description of the evolution of local atomic structures with dissimilar symmetries. Utilizing a sample with a cooling rate of $10^{13}$ K/s, we employ the degree of $d_5$ to label different local atomic regions. In addition to examining the variation in $d_5$ for the corresponding atomic clusters, we also investigate the changes in triangular ($d_3=n_3 / \sum_{i} n_i$), quadrilateral ($d_4=n_4 / \sum_{i} n_i$), and hexagonal faces ($d_6=n_6 / \sum_{i} n_i$) with LTS features during the evolution process (Fig.~\ref{Fig:Fractiond3d4d5d6}). The results further confirm that the thermal stability and earliest evolution of atomic clusters are not correlated with their original local symmetries.

\begin{figure}
	\begin{center}
		\includegraphics[width=0.7\columnwidth]{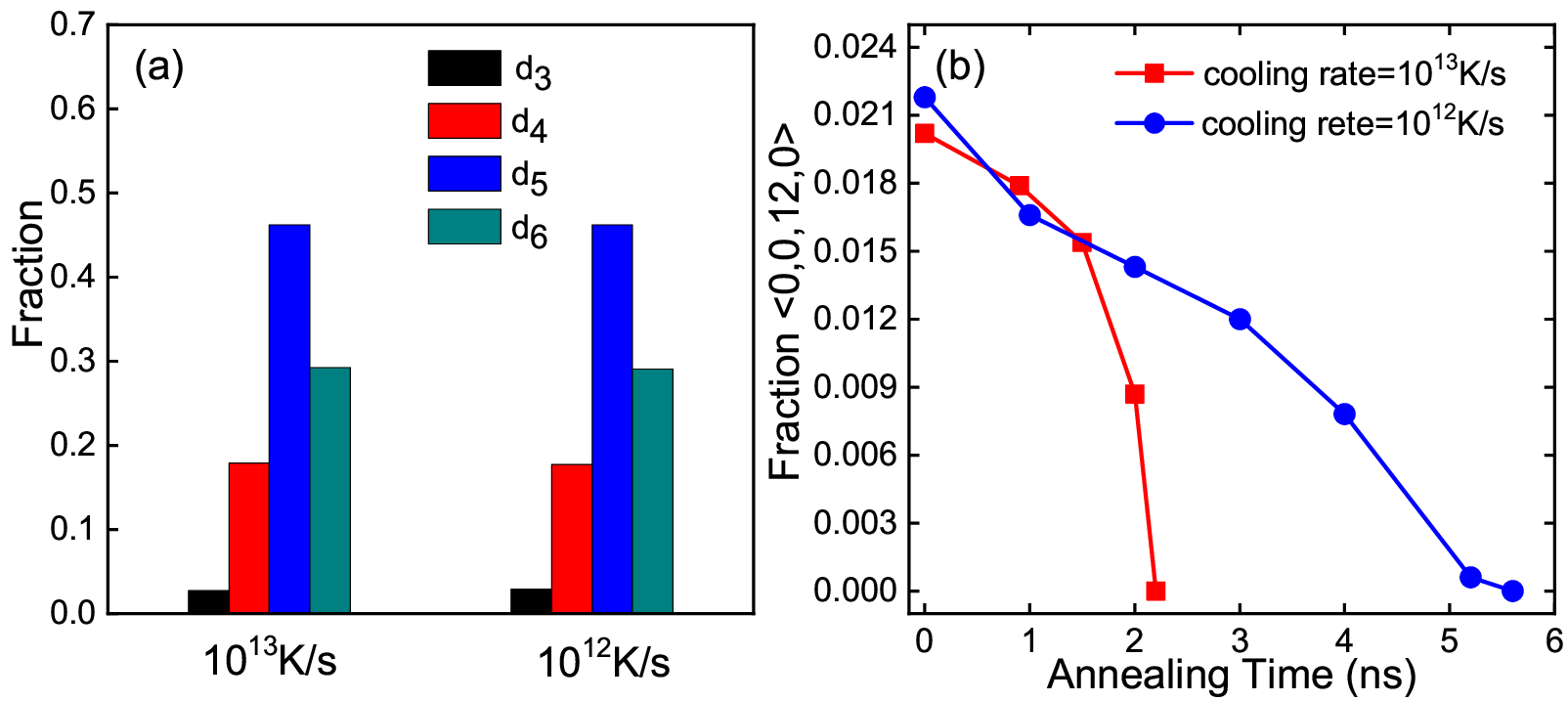}\\[5pt]  
		\caption{\parbox[c]{15.0cm}{\footnotesize{\bf Fig.~7.} (color online)  (a) The fractions of i-edged faces in polyhedra of two samples associated with different cooling rates.  (b) The variation of the fractions of the icosahedral clusters during the isothermal annealing treatment.  }}
		\label{Fig:FractionFraction}
	\end{center}
\end{figure}

After elucidating the initial-stage evolution of the local atomic clusters, our focus now shifts to understanding the factors behind the disparate changes observed in the PE curves which correspond to the differences in thermal stability of samples produced at different cooling rates. Figure~\ref{Fig:FractionFraction}(a) presents the average fraction of pentagonal faces, reflecting the LFFS of the entire system, along with the presence of triangular, quadrilateral, and hexagonal faces exhibiting the LTS features. The result suggests that the different cooling rates we considered here have a negligible impact on the prevalence of both LFFS and LTS in the corresponding samples. As well, the proportions of icosahedral clusters $<0, 0, 12, 0>$ in the two as-quenched samples, as well as their variations during the annealing process were shown in Fig.~\ref{Fig:FractionFraction}(b). The results unequivocally demonstrate that, despite the similar initial fractions of $<0, 0, 12, 0>$ in these two samples generated at different cooling rates, their subsequent evolution over time exhibits substantial disparity. Consequently, we naturally infer that the underlying mechanism accounting for the aforementioned diversity lies in the distinctive packing and correlation of icosahedral clusters within the glass structures. This investigation addresses a crucial question regarding the connectivity and arrangement of local icosahedral clusters, which profoundly influence the thermal stability and crystallization process in these two samples.

\begin{figure}
	\begin{center}
		\includegraphics[width=0.6\columnwidth]{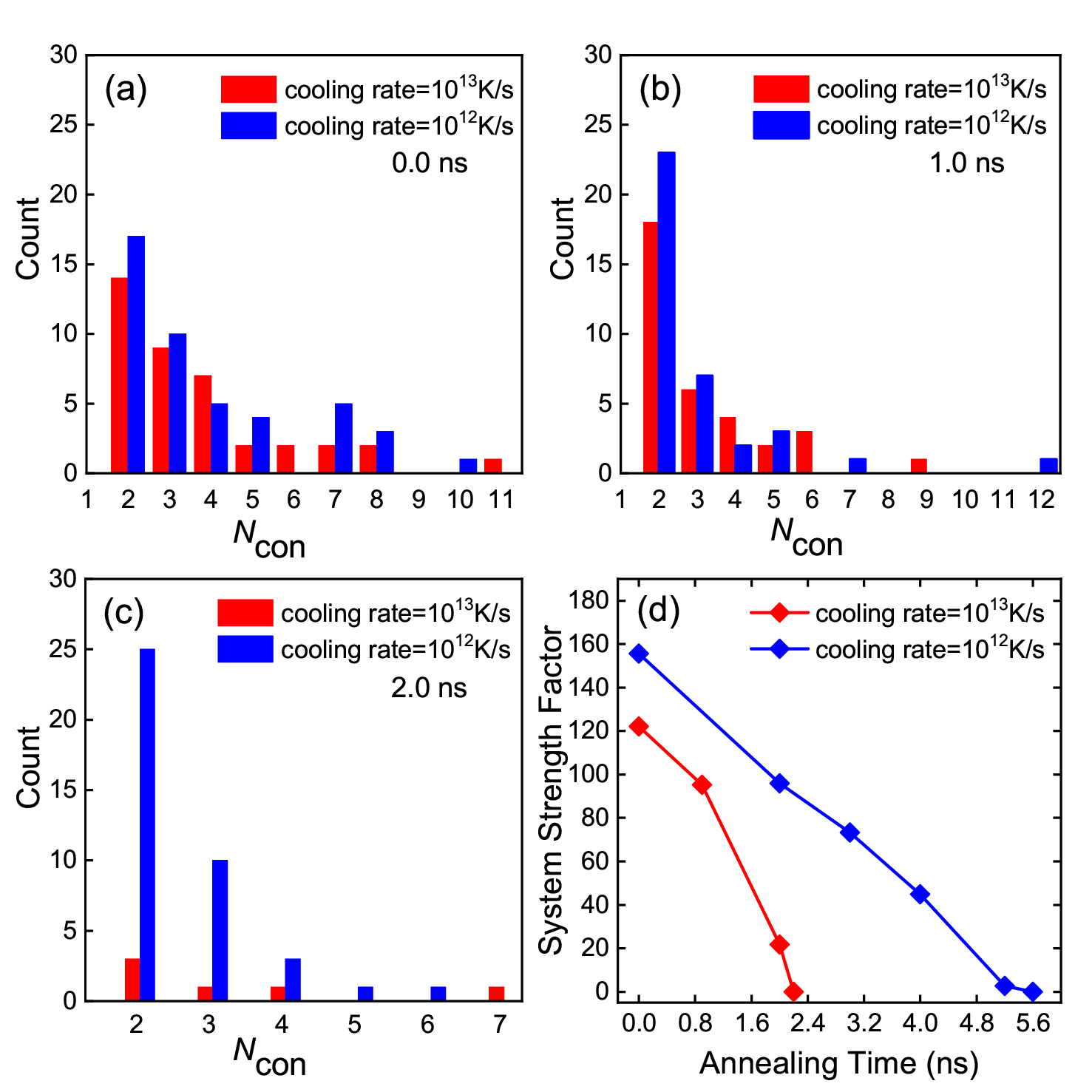}\\[5pt]  
		\caption{\parbox[c]{15.0cm}{\footnotesize{\bf Fig.~8.} (color online)
        The distributions of $N_{con}$ at several time points ((a)$0.0$ ns, (b)$1.0$ ns, (c)$2.0$ ns.) during the isothermal annealing process. (d) The change in system strength factors of two samples with dissimilar cooling rate during the isothermal annealing process.}}
		\label{Fig:CountStrengthFactor}
	\end{center}
\end{figure}

To investigate the network and extended clusters (ECs) formed by the interconnection of the $<0, 0, 12, 0>$ clusters, here we introduced the method of graph theory approach. The central atoms of all icosahedral clusters will be treated as nodes in a network, and if two different central atoms have the same one atom as their common nearest-neighbors, the two represented nodes are considered to be connected. The number of the sharing atoms will be used as the weight of this link. Now the question is transformed into a mathematical problem, obtaining and analyzing the connected components (i.e. the ECs in the glass structures) in a graph. The connectivity number ($N_{con}$) which characterizes the space length scales of ECs is defined as the number of nodes in a connected component in the corresponding graph. Distributions of $N_{con}$ at several time points during the isothermal annealing process have been shown in Fig.~\ref{Fig:CountStrengthFactor}. It can be seen that the icosahedral clusters in the samples generated under a lower cooling rate exhibit stronger spatial correlations, and they tend to be connected with each other. The results indicate that the low cooling rate will enhance the spatial correlations of the icosahedral clusters and consequently increase the stability of the ECs network in the sample, which would be the key factor of dynamic slowing down to prevent the crystallization. In order to make different dimensional ECs have comparability, we define strength factor for an EC as $\sum w_i / N_{con}$, where $N_{con}$ is the number of nodes in an EC, and $w_i$ is the sum of weight of all links in this connected component abstracted for the corresponding EC. The global strength factor of the network representing the atomic structure will be defined as the sum of the strength factors over all ECs. The outcome implies that the two adopted cooling rates lead to small difference (about $7.3\%$, see Fig.~\ref{Fig:FractionFraction}) in the quantity of icosahedral clusters in two samples, but large influence (about $27.4\%$, see Fig.~\ref{Fig:CountStrengthFactor}) on the global strength factor of the samples. The results demonstrate once more that the thermal stability against crystallization do not depend on the total quantity of the icosahedral clusters, but on the degree of connectivity among them in our cases. As well, the variation of the system strength factor also has been given in Fig.~\ref{Fig:CountStrengthFactor}. It can be seen that systems subjected to different cooling rates exhibit distinct evolving behavior during the crystallization process, which indicates that the system strength factor is a good indicator for the thermal stability of the MGs and strongly correlated with the evolutionary time-scales in the crystallization process of them.
And this also can be used to rationalize what we found in Fig.~\ref{Fig:PE}(b), the bimodal feature, by the fact that in the slow-quenched samples the stochastic nucleus's development and the subsequent crystal growth will be greatly influenced by how close it is to a compact extended icosahedral-clusters, where the connectivity strength is relatively high.

The crystallization is a complex physical process consisting of two major events, nucleation and crystal growth which are driven by thermodynamic, chemical, dynamic, as well as local structure properties. It has been well documented in the literatures that the presence of icosahedral clusters and of connected icosahedra plays an important role on the dynamic slowing-down of the metallic liquids during supercooling\ucite{Cheng2011Atomic,Wu2013Correlation}. Therefore, what we found here can be attributed to the fact that a more compact icosahedral network will stabilize the dynamics of the liquids a bit and then hinders the crystal growth process. At the same time, even more interesting and complicated, the present of the icosahedra network possessing lower formation energy\ucite{Wu2016critical} could also increase the free energy barrier of the nucleation process in the view of it makes the interface between the crystal nuclei and the surrounding medium a bit rough (effectively large interface area), which leads to a potentially more interface free energy cost and then against the crystallization\ucite{Sun2013effects,Desgranges2018unusual}. In other words, the increase of crystalline embryos and the decrease of icosahedral connectivity during annealing seem to imply that there is a mutual constraint between the two factors. On the one hand, with the growth of crystalline embryos, the connectivity of icosahedra is damaged, which drives the whole crystallization process; on the other hand, the connectivity of icosahedral clusters inhibits the formation of crystalline embryos. Especially when a much slower cooling rate was adopted, the crystal-like short-range orders may start to organize to form some extended structures, which would definitely influence the following crystallization process when the sample was annealed at some temperatures. The interplay between the embryos of crystallization and the connectivity of icosahedra on the thermal stability of the liquid will become even more complicated at that time\ucite{Desgranges2018unusual,Desgranges2019can}, and the mechanism of the interaction between these two factors in the metallic glass-formimng systems is a thought-provoking issue that deserves an in-depth investigation in the future.

\section{Conclusion}
In this study, we provided a refined description of the evolutionary details of the atomic structures with different local symmetries in the crystallization process of a metallic glass. At the very beginning of the devitrifaction, the results suggest that the thermal stability and the earliest atomic structure evolution have no correlation with their short-range order (SRO) characters, which leads to a novel phenomenon of local symmetry convergence being found. We propose that this could be the key mechanism for the nucleation of high-density nanocrystals at the beginning of the crystallization process of MGs under the deep supercooled conditions\ucite{Wang2011Atomic,Liu2008Atomistic,Schroers1999Continuous,Schroers2000Crystallization,Pauly2010Criteria}.
Furthermore, it has been clarified that in our system the thermal stability against crystallization does not depend on the total amount of the icosahedral SRO, but on the degree of global connectivity among them, which is a result might promote future studies of the theory for nucleation and growth of crystals in the disordered systems. At last, it is worth to note that the global strength factor which quantifies the degree of global connectivity of the icosahedral networks would also be an interesting alternative to investigate other relevant physical processes of the MGs\ucite{Li2017Local,Wu2018Streched}.



\addcontentsline{toc}{chapter}{Acknowledgment}
\section*{Acknowledgment}
Z.W.W. thanks the members of the ``Beijing Metallic Glass Club'' for the long-term fruitful discussions. This work was supported by the National Natural Science Foundation of China (Grant Nos. 52031016 and 11804027). The work was carried out at National Supercomputer Center in Tianjin, and the calculations were performed on Tianhe new generation supercomputer.

\addcontentsline{toc}{chapter}{References}
\bibliographystyle{iopart-num} 
\bibliography{crystallization}

\end{CJK*}  
\end{document}